# A theoretical study on the mechanisms of formation of primal carbon clusters and nanoparticles in space

Dobromir A. Kalchevski[1], Dimitar V. Trifonov[1], Stefan K. Kolev[1], Valentin N. Popov[2], Hristiyan A. Aleksandrov[3], Teodor I. Milenov[1]

[1] "Acad. E. Djakov" Institute of Electronics, Bulgarian Academy of Sciences, Sofia, Bulgaria
[2] Faculty of Physics, University of Sofia, 5 J. Bourchier Blvd., 1164 Sofia, Bulgaria
[3] Faculty of Chemistry and Pharmacy, University of Sofia, 1 J. Bourchier Blvd., 1164 Sofia, Bulgaria
*Corresponding Author: Stefan K. Kolev, ORCID: 0000-0002-9766-7826, e-mail: skkolev@ie.bas.bg

**Abstract**
We present a computational study of assembling carbon clusters and nanophases in space from carbon aggregations. Geometry optimizations and Density-functional-based tight-binding (SCC-DFTB) dynamics methods are employed to predict carbon clusters, their time evolution, and their stability. The initial density of the aggregates is found to be of primary importance for the structural properties of the clusters. Aggregates with sufficiently low initial density yield clusters with approximately equal prevalence of sp and $sp^2$ hybridized states and almost missing $sp^3$ ones. The increase in the initial density results in $sp^2$-dominant molecules that resemble the carbon skeleton of polycyclic aromatic hydrocarbons (PAHs). Larger initial aggregations with tetrahedral interatomic orientation result in $sp^2$-dominant multi-dimensional polymers. Such materials are highly porous and resemble axially bound nanotubes. Some resultant clusters resemble fullerene building blocks. Spheroid nanoparticles resembling improper fullerenes are predicted by metadynamics, aimed at inter-fragment coupling reactions. One such structure has the lowest binding energy per atom among the studies molecules. All zero-dimensional forms, obtained by the simulations, conform to the experimentally detected types of molecules in space. The theoretical IR spectrum of the clusters closely resembles that of fullerene C70 and therefore such imperfect structures may be mistaken for known fullerenes in experimental infrared (IR) telescope studies.

1. Introduction

More than any other element, carbon has the innate ability to catenate into diverse, structurally rich chains of arbitrary length. Unsurprisingly, the possible organic molecules outnumber the inorganic ones by two orders of magnitude. Property-wise, carbon-based chemistry has no rival. In fact, for over a decade, it has been expanding in experimental and industrial fields, typically designated for inorganic materials[1]. Form variety and reactivity in mild conditions enable carbon as the fundamental building block of biological systems - the most complex of the known structures. A basic fundamental question is that of the formation and reformation of carbon phases in the vastness and harsh conditions of space. Astrophysics has been answering questions about the origin and distribution of this intriguing element in the Universe. However, it is up to chemistry to answer what would be the very first structures carbon atoms form upon their first meeting in conditions enabling reactivity towards complex molecules. This is the scope of the current scientific study. The origin of carbon in the Universe is the triple-alpha fusion process in stellar cores. Star collapse and supernova explosions disperse the atoms in the interstellar and intergalactic space. While circumstellar envelopes are considered the largest reactors for carbon chemistry in space[2,3], pathways to complex carbon molecules also exist in interstellar clouds[4,5]. Carbonaceous compounds are also found in distant galaxies[6,7] and protoplanetary nebulae[8-10]. Although the detection of carbon in exoplanetary atmospheres has many obstacles, and some claims are being debated[11], some are considered reliable[12-14]. Surprisingly, Polycyclic Aromatic Hydrocarbons (PAH) were found even in the cold, dark Taurus Molecular Cloud 1 (TMC-1), at a temperature of 10 K, far away from any evolved star[15]. In our solar system, carbon has been registered in the atmospheres of planets and moons[16-18], and even in comets[19-21].

To date, of 256 molecules detected in the cosmos, the existence of 204 is proven, and most of them contain carbon[22,23]. The size of the structures ranges from 2 to 70 atoms[23]. The first carbon chemical to be discovered in space is CH (methylidyne)[24], CN (cyan radical)[25], CS (carbon monosulfide)[26] and $C_2$ [27] are other examples of such small molecules. Clusters of large carbon content can give us an insight into the preferred structures in cosmic environments consisting primarily of the element of life: $l$-$C_3H^+$ (cyclopropynylidynium cation)[28], HCCO (ketenyl radical)[29], $C_4H$ (butadiynyl radical)[30], $C_5$ (pentacarbon)[31], $H_2CCC$ (propadienylidene)[32], $C_5N$ (cyanobutadiynyl radical)[33], $C_6H$ (hexatriynyl radical)[34], $HC_7N$ (cyanotriacetylene)[35], $C_8H$ (octatriynyl radical)[36], $HC_9N$ (cyanotetraacetylene)[37], $C_6H_6$ (benzene)[38] and $C_6H_5CN$ (benzonitrile)[39]. Intriguing small to medium-sized cyclic molecules have been detected in TMC-1: $C_3H_2$ (cyclopropenylidene)[40], $C_5H_6$ (cyclopentadiene)[40], $C_5H_5CN$ (cyanocyclopentadiene)[41], $C_5H_5CCH$ (ethynyl cyclopentadiene)[42] and $o$-$C_6H_4$ (ortho-benzyne)[43]. Even more intriguing are the polycyclic species detected in TMC-1: $C_9H_8$ (indene)[40] and two isomers of $C_{11}H_7N$ (1- and 2-cyanonaphthalene)[15]. All experimentally known allotropes of carbon (diamond, graphite, fullerenes) have been detected in space[44]. Of special interest are the three fullerene structures discovered so far: $C_{60}$ (buckminsterfullerene)[45], $C_{60}^+$ [46], and $C_{70}$ (rugbyballene)[45].

Since experimentally simulating the exact harsh environment of space and specifics in the events of original carbon atom aggregation poses many obstacles, we turn to theoretical chemistry to answer the postulated question of initial carbon clusters. Quantum chemistry has a decades-proven track record of modeling reactivity and explaining the relationships between structure and properties. Carbon-wise computerized research has been successfully employed in anything from predicting and modulating physical and chemical properties of novel theoretical allotropes[47-49], through phase transitions of solid-state materials[50], to complete ground state and excited state reactivity of organic molecules, and reactions mechanism characterization[51].

Models of carbon allotropes, such as D-carbon[52], modulated T-carbon[53], and ultra-hard rhombohedral carbon[54] have been employed to yield total energies, pressure stability, and band gaps. Datta et al.[55] have investigated the effects of the introduction of Stone-Wales defects on the metallicity of carbon nanotubes. Marchant et al.[56] have explored the configurational space of elemental carbon, to determine macroscopic properties, melting transitions, and thermodynamically stable structures in a wide range of pressures. A simple simulation code for the formation of planar hydrocarbon clusters has been proposed[57]. The IR spectra of a generated population of C60 isomers has been compared to that of buckminsterfullerene, to discover that the plateau in the 6 – 9 μm region can only be observed from closed-cage structures[58]. The optical spectra of generations of $C_{n=24,42,60}$ clusters have been assigned over structural characteristics, such as asphericity and hybridization, in order to provide an alternative explanation for a UV bump in the interstellar medium extinction curve of galaxies[59]. The possibility of the formation of hydrogenated amorphous carbon nanoparticles from PAHs has been proposed[60]. Amorphous carbon at low densities has been evaluated in terms of hybridization prevalence, density of vibrational states, and localized vibrational modes[61] – this appears to be the only previous research using disordered atom arrangements as starting structures, instead of accepted experimental or theoretical carbon formations. Another theoretical study investigates the relationship between density and structural characteristics of amorphous carbon, such as hybridization prevalence and interatomic orientation[62]. High-level multi-configurational theory has revealed that the cyclic isomer of $C_3^+$ is 5.2 kcal/mol more stable than the linear[63]. Takahashi[64] has predicted certain small-sized $C_n$ (n=2-8) interstellar carbon species, focusing on stability and proposing mechanisms of formation. The 3-atomic ring systems have been found more stable than the linear isomers. A theoretical model of a mechanism has been proposed for the formation of carbon nanotubes in space[65]. Two major approaches exist, as propositions on the formation of fullerenes in space. The "top-down" approach starts with the highly ordered graphene[66-68], while the "bottom-up" approach initiates with smaller molecules[69-72]. A comment article from 2022 stresses the open question of formation and destruction mechanisms for carbon molecules in space, among problems of spectral band assignment[73]. To the present day, the theoretical investigation of carbon clusterization is incomplete hence the importance of the current research lies in the fulfilment of the task.

This study is focused on the theoretical investigation and characterization of the paths of formation of original carbon clusters in space. The characterization of the resulting systems involves the role of orientation, size, and density of the initial, pre-reaction atomic aggregations. The cluster description includes the prevalence of structural elements and hybridization. The stabilization of the systems was conducted with DFTB molecular dynamics. The resulting molecules, resembling building blocks for closed-cage structures, were assembled into semi-spherical nanoparticles with DFTB metadynamics. To our knowledge, this state-of-the-art method for reaction mechanism research has previously not been employed for carbon clusterization. The theoretical IR analysis indicates that even semi-spherical nanoparticles, assembled from disordered initial atomic formations, can be misidentified by experimental means as known fullerenes.

## 2. Methods

All calculations are performed with the CP2K/Quickstep package[74,75]. The SCF optimizations are completed by the Self-Consistent Charge Density Functional Based Tight Binding (SCC-DFTB / DFTB2) method[76]. An efficient *a posteriori* treatment for dispersion interaction is employed[77].

The DFTB method is an approximation to Density Functional Theory (DFT) in which the Kohn-Sham (KS) equations are transformed to a form of tight binding ones[78] related to the Harris functional[79]. The here-used second-order expansion of the KS equations enables a transparent, parameter-free generalized Hamiltonian matrix. Its elements are modified by a self-consistent redistribution of Mulliken charges (SCC)[80]. The KS energy additionally includes a Coulomb interaction between charge fluctuations. The accuracy of DFTB with SCC expansion is comparable to that of DFT methods and higher levels of ab initio theory for various properties of single molecules, solutions, and solid-state materials. The method yields satisfactory geometries and total energies[81-83]. DFTB produces good charge distribution, binding energies, and vibrational frequencies of charged solvated species[84]. Activation energies in organic chemistry also conform to higher levels of theory[82], enabling the study of reaction mechanisms. It has been found that the method yields excellent geometries and energetics for pure carbon species, such as fullerenes ranging from C20 to C86[85]. At the same time, the method is orders of magnitude less computationally intensive than DFT. The binding energies per C atom for all $C_n$ clusters formed are calculated as $E_b = [E(C_n) - n \times E(C)]/n$, where $E(C_n)$ and $E(C)$ are the energies of the corresponding cluster and an isolated carbon atom, respectively, while n is the number of C atoms in the cluster. With this definition, a lower (more negative) $E_b$ value corresponds to a more stable $C_n$ cluster.

All simulations are performed with Born-Oppenheimer Molecular Dynamics (BOMD)[86]. The method of metadynamics (MTD) is chosen to model the formation of nanoparticles[87,88]. Metadynamics is a state-of-the-art simulation method in which the processes are guided to model a chosen chemical reaction. Collective variables (colvars, CVs) are defined over molecular degrees of freedom to bias the system towards selected changes. Penalty potentials (hills) are periodically spawned for current values in the CV space to raise the free energy and reach unexamined geometries. When a TS is crossed over, the study of the new minimum begins, once again starting at the bottom of the energy pit. Reversing the bias potential peaks gives us the relative stability of each geometry – the free energy surface of the reaction. With the raise in energy, unguided changes are free to occur, giving a realistic insight into the studied processes.

Metadynamics simulations are carried out in the NVT ensemble. The thermostat is canonical sampling through a velocity rescaling (CSVR)[89,90], set for a temperature of 400 K. The timestep is 1 fs. The height of the Gaussian penalty potentials is 1.255 kcal/mol. The scale factor (Gaussian width) for each collective variable is 0.2. Hills are spawned every 50 fs. All walls are of quadratic type with a potential constant of 20 kcal/mol. The temperature tolerance is always set to 50 K. The shortest intermolecular distances are above 3.4 Å in all initial (zeroth) steps. Each MTD run is preceded by geometry optimization of the systems.

## 3. Results and discussion

*1 Structural tendencies for carbon clusters in space*

The structural prevalence of carbon clusters in space is studied by a theoretical examination of different initial atom aggregations. Each such formation undergoes two computational steps to model the natural appearance of primal carbon molecules. The first step is geometry optimization. The second step is BOMD, enabling the spontaneous reformation into a more stable form. Dynamical simulations are carried out in NVT ensemble if the goal is a lone cluster, or in NPT to enable polymer formation. The NPT ensemble tailors the cell dimensions to those of the modeled system and allows intercellular covalent bonding into multidimensional systems. Atom hybridization is determined on the basis of number of neighboring atoms, orbital orientation, bond lengths and three-point angles. The binding energies per C atom ($E_b$) are estimated after geometry optimization of the final dynamics structures.

*1.1 Carbon Cluster I*

The first studied system consists of 25 carbon atoms, randomly positioned in 3D space and centered in a 15.0 × 15.0 × 15.0 Å cell (Fig 1a). The initial density is 1.25 g/cm$^3$. Geometry optimization yields a cluster with approximately 2D geometry, resembling the C-skeleton of a polycyclic aromatic hydrocarbon (PAH). According to molecular topology and bond lengths sp$^2$ hybridization is around three times more dominant than sp$^3$. There are no atoms in the sp state. The molecule consists of networks of conjugated double bonds. To ensure stability of the final structure (Fig 1c) 20 ps of DFTB molecular dynamics in NVT ensemble is carried out. A single reaction with a minor structural significance occurs: the breaking of a C - C bond. The product, named CCI, conforms to types of cyclic molecules, experimentally detected in space[38,39,41-43], especially the polycyclic indene[40]. The $E_b$ is estimated to -7.52 eV. Additionally, CCI has a resemblance to a fullerene building block and is used for the assembly of a closed-cage structure (later in the article).

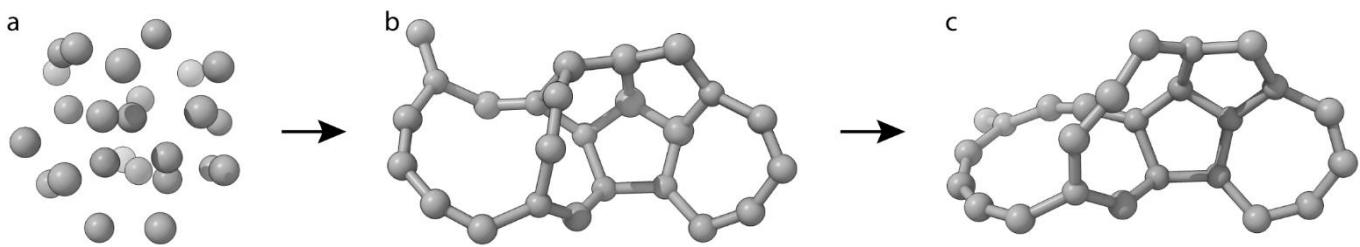

**Figure 1**: Geometries of the initial carbon atom arrangement (a), the structure after geometry optimization (b) and the final structure (CCI) after DFTB molecular dynamics (c). The carbon atoms are colored gray.

*1.2 Carbon Cluster II*

The second studied system consists of 22 carbon atoms, arranged in a tetrahedral pattern and centered in a 15.0 × 15.0 × 15.0 Å cell (Fig 2a). The initial density is 1.37 g/cm$^3$. Geometry optimization results in a tangled polycyclic 3D network of rings with varying sizes (Fig 2b). Molecular topology and bond lengths state that the atom hybridization has the following prevalence: sp$^2$ > sp$^3$ > sp. The carbon atoms in the sp$^2$ state are almost twice as those in sp$^3$. During 20 ps of DFTB molecular dynamics, in the NVT ensemble, ring-opening and ring-closing reactions yield significant structural rearrangements. The resultant molecule (CCIITMP) resembles a C-skeleton of PAH and its geometry (Fig 2c) is close to planar. The predominant hybridization of the carbon atoms remains sp$^2$. CCIITMP somewhat resembles a fullerene fragment. Metadynamics is employed to reform the structure, into a building block for a closed-cage molecule, closer to experimentally known fullerenes. Two distance colvars (Fig 3a) are set. In the course of the simulation, unbiased chemical processes (Fig 3a) result in a cluster of satisfactory geometry (named CCII, Fig 3b). Only one of the anticipating covalent changes according to the CVs was necessary. The free energy profile of the occurred guided reaction is in Fig 4. The process is exothermic, with a product quite more stable than the reagent: the forward barrier is 9 kcal/mol, while the reverse is 32 kcal/mol. The $E_b$ of CCII is estimated to -7.63 eV. This molecule is a single conjugated system of double and triple bonds. The cluster is used for the assembly of a closed-cage, fullerene-like structure (later in the article).

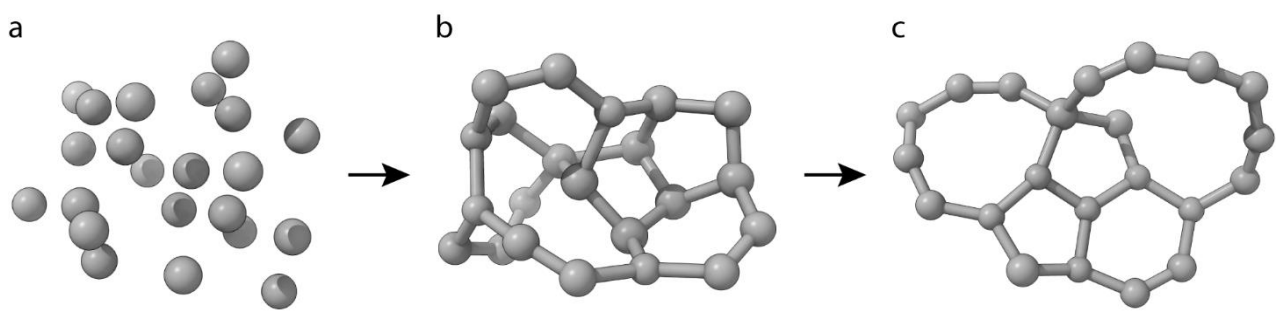

**Figure 2**: Geometries of the initial carbon atom arrangement (a), the structures after geometry optimization (b) and the cluster (CCIITMP) after DFTB molecular dynamics (c).

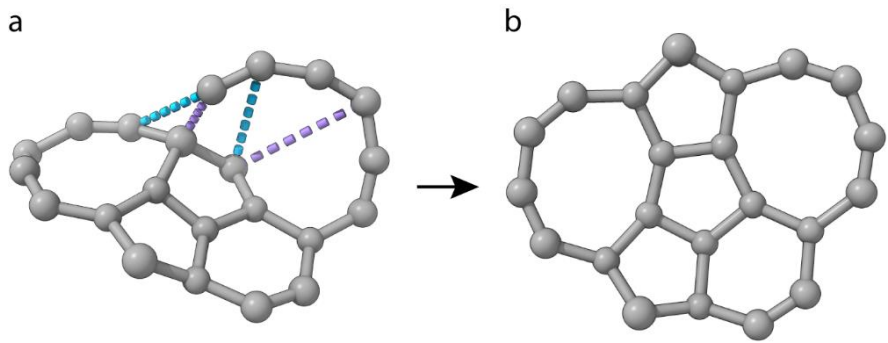

**Figure 3**: Distance CVs (in violet) and distances of additional, unbiased chemical changes (in blue) in the MTD of reforming CCIITMP (a) and the final result – CCII (b).

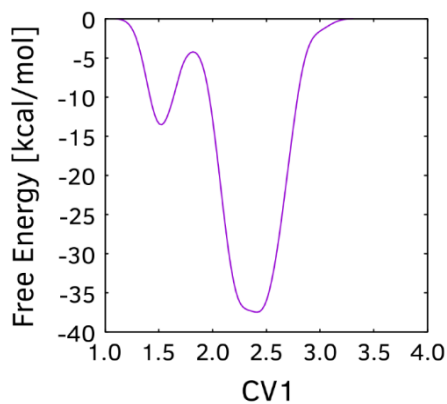

**Figure 4**: Free energy profile of the guided reaction in the chemical transformation of CCIITMP to CCII.

*1.3 Carbon Cluster III*

The initial formation consists of 20 carbon atoms, arranged into a somewhat symmetrical geometry (Fig 5a). The system is centered in a 16.0 × 16.0 × 16.0 Å cell. The mean interatomic distance is approximately twice that of the aggregation to result in CCI. The initial density is 0.57 g/cm$^3$. Geometry optimization results in a structure, chemically identical to the one in Fig 5b. According to topology and bond lengths half of the atoms are in sp hybridization and the other half are in sp$^2$. There are no carbons in the sp$^3$ state. Two rings are formed and both are 3-atomic. The structure is mostly branched linear. DFTB molecular dynamics in the NVT ensemble yields no chemical reactions in 20 ps (Fig 5b). $E_b$ is found to be -7.15 eV. The linear fragments of the structure correspond to experimentally detected molecules in space[31,33-37]. According to our results, it is expected that such molecules appear when the original density of the carbon aggregation is low. The spontaneous formation of 3-atomic rings conforms to a previous theoretical conclusion, that such rings are energetically preferred to linear structures when it comes to chains of up to 7 carbon atoms[64].

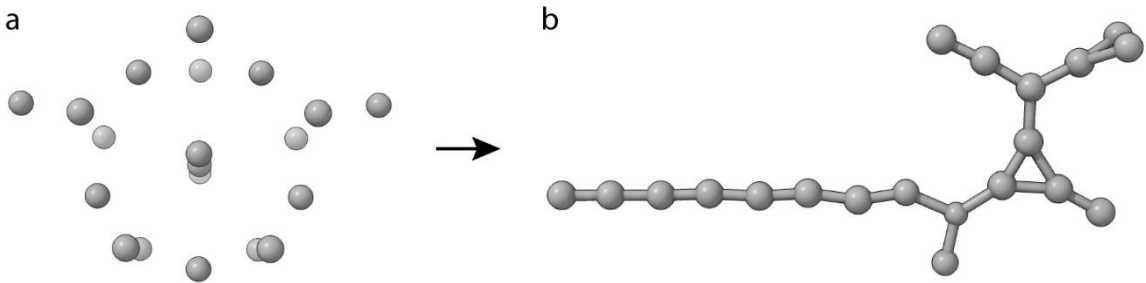

**Figure 5**: Geometries of the initial carbon atom arrangement (a) and the final structure (CCIII) after DFTB molecular dynamics (b). The carbon atoms are colored gray.

*1.4 Carbon Cluster IV*

A larger cluster of approximately double the previous sizes is simulated in a supercell with 3D Periodic Boundary Conditions (PBC). The initial geometry is 40 carbon atoms in a tetrahedral arrangement (Fig 6a), centered in a 13.5 × 16.5 × 13.5 Å cell. The initial density is higher than that of CCIII and close to the densities of CCI and CCII (1.6 g/cm$^3$). Geometry optimization results in the structure in Fig 6b. The structure is tangled polycyclic. The prevalence of hybridization is sp$^3$ > sp$^2$ > sp. There are almost as many atoms in the sp$^2$ state as in sp$^3$. In our entire research, this is the only structure in which the dominant hybridization is sp$^3$.

DFTB molecular dynamics is carried out for 20 ps in an NPT ensemble. A supercell with initial dimensions of 9.0 × 12.0 × 10.0 Å is used with 3D PBC. After many ring-opening and ring-closing reactions, the structure is significantly reformed into a 2D polymer. The elementary cell is a complex network of fused rings, arranged in a somewhat tubular formation (Fig 6c). The single cell dimensions are 8.02 × 10.69 × 8.91 Å. The predominant hybridization is once again sp$^2$. $E_b$ is estimated to -8.17 eV. The polymer somewhat resembles a network of laterally bound carbon nanotubes with a weak structure definition (Fig 7).

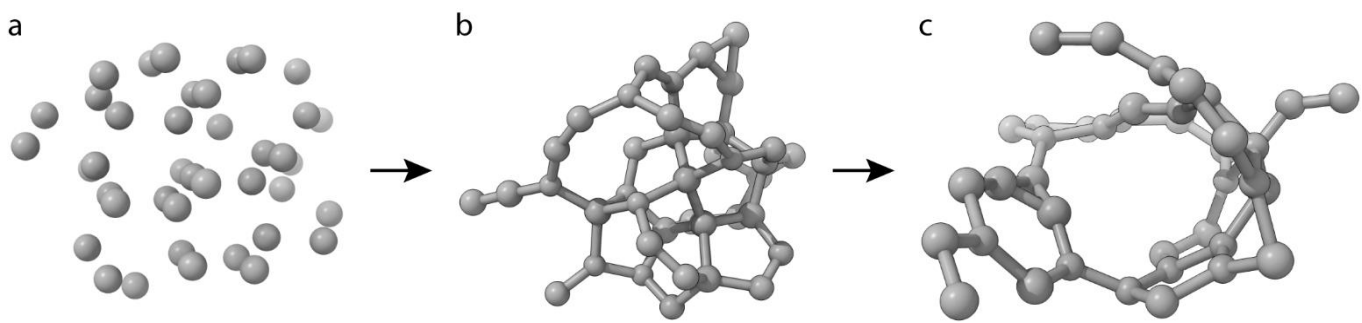

**Figure 6**: Geometries of the initial carbon atom arrangement (a), the structure after geometry optimization (b) and the final cluster (CCIV) after DFTB molecular dynamics.

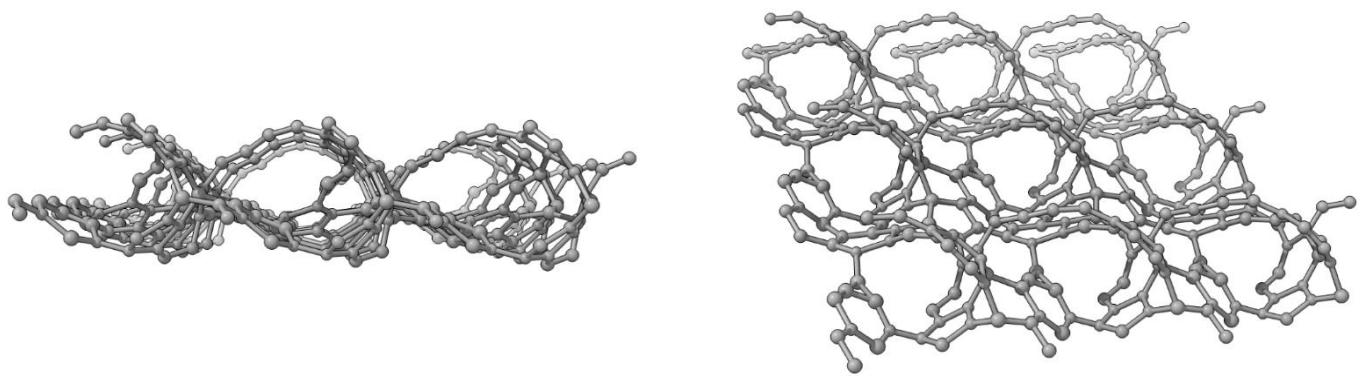

**Figure 7**: 3×1×3 multicell views of the 2D polymer CCIV.

The event of carbon atoms congregating in vacuo results in various formations, depending on the number of atoms and the initial density. Lone carbon aggregations of 20 to 25 atoms with sufficiently high-density form $sp^2$-dominant clusters representing the C-skeleton of PAHs. The result is independent of the initial interatomic orientation (for example random or tetrahedral). Such clusters have a resemblance to fullerene building blocks. Similar-sized aggregations at sufficiently reduced density result in branched molecules with far more prominent sp hybridization, a very similar degree of $sp^2$ hybridization, no $sp^3$ hybridization, and very few rings of low atomic number. If the PBC model has varying cell dimensions, tailored by the cluster size, $sp^2$-dominant multi-dimensional polymers form at the initial higher density. Such materials exhibit porous structure and somewhat resemble laterally-bound nanotubes. It is possible that at really low temperatures and/or short periods $sp^3$-dominant tangled-polycyclic clusters occur in cases of large, pre-reaction carbon aggregations, as evident from the pre-dynamics optimization of CCIV. DFTB molecular dynamics runs sometimes lead to stabilizing reforming of the structures. The procedure tends to increase the number of atoms in $sp^2$ hybridization and decrease the number in $sp^3$ hybridization, if originally present. Hence, we also expect such tendency to guide spontaneous transformations in nature. Structures arising from initial aggregations with higher density tend to be more stable, according to binding energy per C atom. Enabling polymerization, with theoretically unlimited number of atoms, takes the tendency further. Spontaneous covalent coupling between smaller clusters is expected, in the event of their gathering. All 0-dimensional clusters conform to experimentally detected types of carbon molecules in space.

*2 The formation of spherical carbon nanoparticles*

The binding of clusters into nanoparticles is simulated with metadynamics.

*2.1 Modeling the synthesis of nanoparticle C71SC*

Metadynamics is employed in an attempt to react identical CCI instances into a spherical nanoparticle. At first 2 CCI clusters (Fig 8a) are bound to a half-sphere. The fragment-coupling reactions are guided with four C – C distance CVs. Three of the interfragment bonds appear almost without a barrier (< 2 kcal/mol, Fig 9), the fourth requires 11 kcal/mol due to the necessity for a small whole-system twist (Fig 9). In the reactions guided with CV2 a neighboring atom participates in the bond formation instead of a targeted atom and an artifact appears in the free energy profile – the covalent interatomic distance minimum has an unexpectedly large value of 2.5 Å. Since penalty potentials are automatically generated at even intervals perhaps the negligible barriers found are simply an artifact of the time required for the clusters to bind at product bond lengths. A product is formed in less than 1 ps. The resulting cluster, named simply 2xCCI (Fig 8b), has 34 atoms in its round fragment and 16 atoms in side chains. The product free energy pit is not completely studied, but a reverse barrier of 37 kcal/mol at the end of the simulation guarantees that the product is an order of magnitude energetically more stable than the reagents.

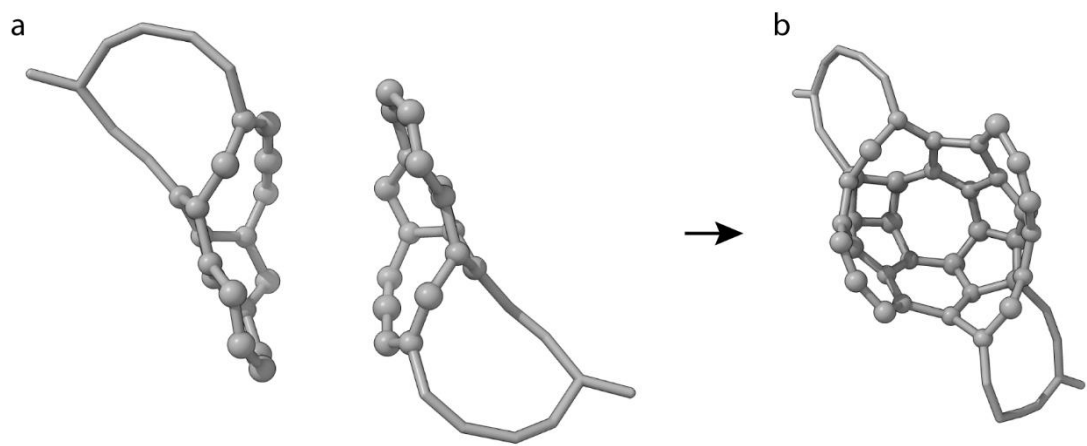

**Figure 8**. Geometries of the initial arrangement of the two CCI clusters (a) and the final structure (2×CCI) after DFTB metadynamics.

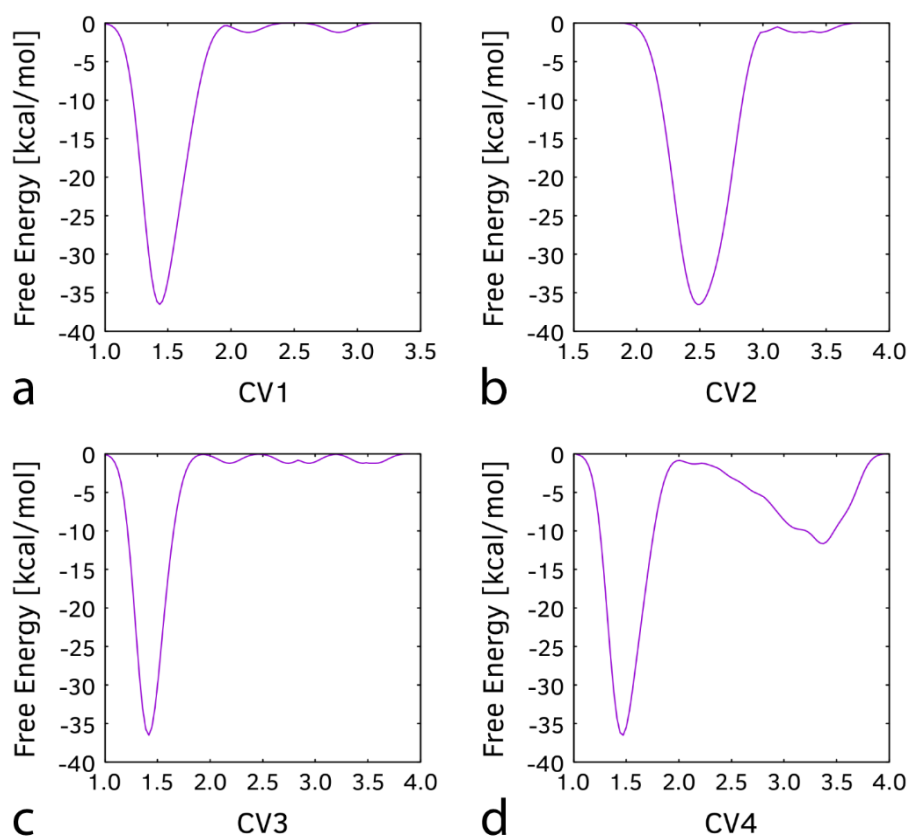

**Figure 9**: Free energy profiles of the guided four of the coupling reactions in the formation of 2×CCI. The product minima are not completely explored.

Two 2×CCI (Fig 10a) react with each other in another MTD step to a nanoparticle (NP) with an entirely spherical fragment. The fragment-coupling reactions are guided with four C – C distance CVs. Three of the interfragment bonds appear in an almost barrierless fashion and a product forms in less than 1 ps. The small barriers found (< 2 kcal/mol, Fig 11) are once again probably an artifact of the time required for the fragments to bind at product bond lengths. After the first three of the bonds are formed the positions of the two carbon atoms in CV2 are less appropriate for coupling which leads to a single barrier of 11 kcal/mol (Fig 11b). Further fragment-binding bonds are formed without CV bias to a structure with 71 atoms with spherical shape (C71SC, Fig 10b). The barrier for the reverse process is found to be 37 kcal/mol which shows the notable stability of the C71SC structure.

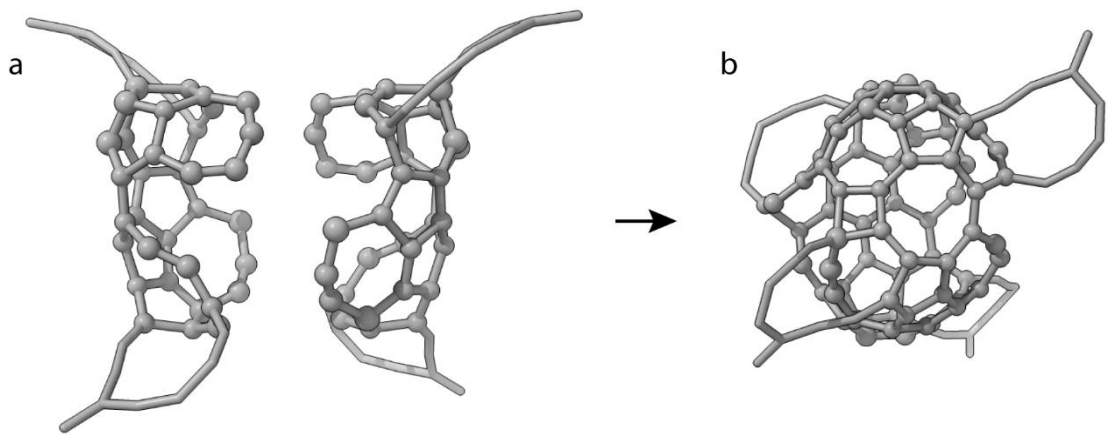

**Figure 10.** Geometries of the initial arrangement of the two clusters (a) and the final structure (C71SC) after DFTB metadynamics.

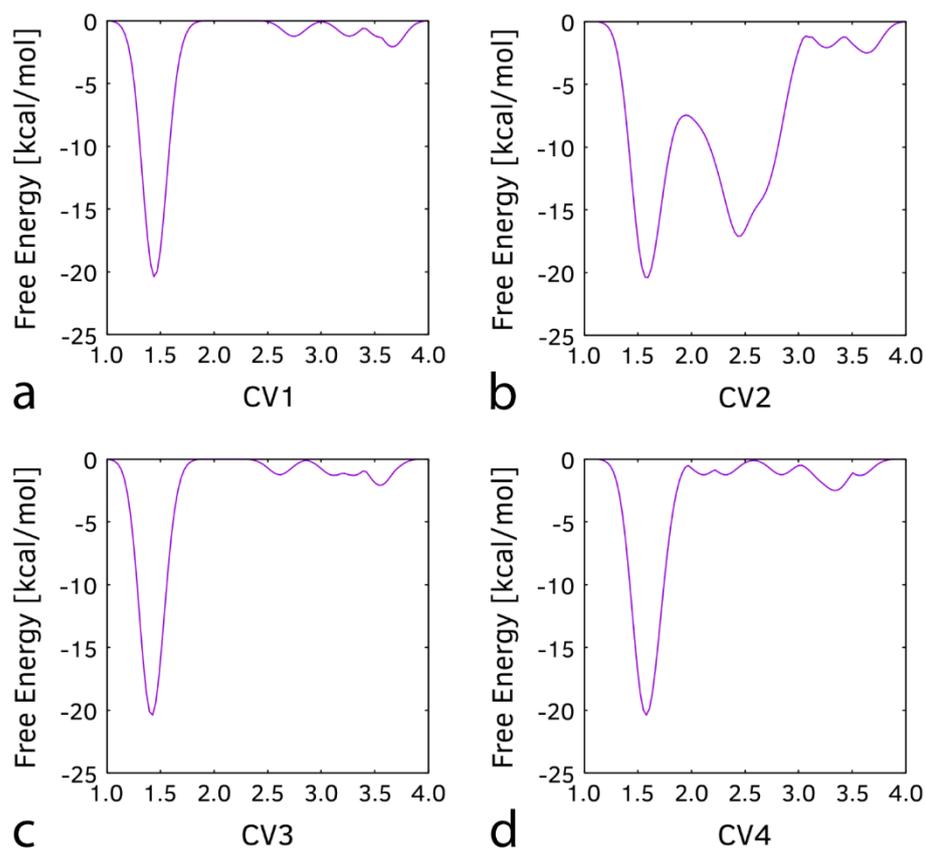

**Figure 11**: Free energy profiles of the guided four of the coupling reactions in the formation of C71SC. The product minima are not completely explored.

*2.1.2 Structural improvements of C71SC*

Improvements in the structure of C71SC are carried out with MTD to achieve a more symmetrical shape. A few gaps are to be closed by intramolecular bond formation. Four C – C distance CVs are used and the barriers found are: 0.5, 2.5, 7, and 24 kcal/mol (Fig 12). The product-free energy pit is once again not entirely explored but the barrier of 32 kcal/mol at the end of the simulation for the reverse process means that the product is significantly more stable than the reagent and gap-reopening reactions are very unlikely. The final fullerene-like nanoparticle is C72SC and represents a C72[4,5,6,7,8,10]fullerene with side chains (Fig 13). $E_b$ is found to be -8.01 eV. The spherical core is not as symmetrical as fullerenes C60 or C70, but considering the process started with 25 randomly positioned atoms, the final result is significant.

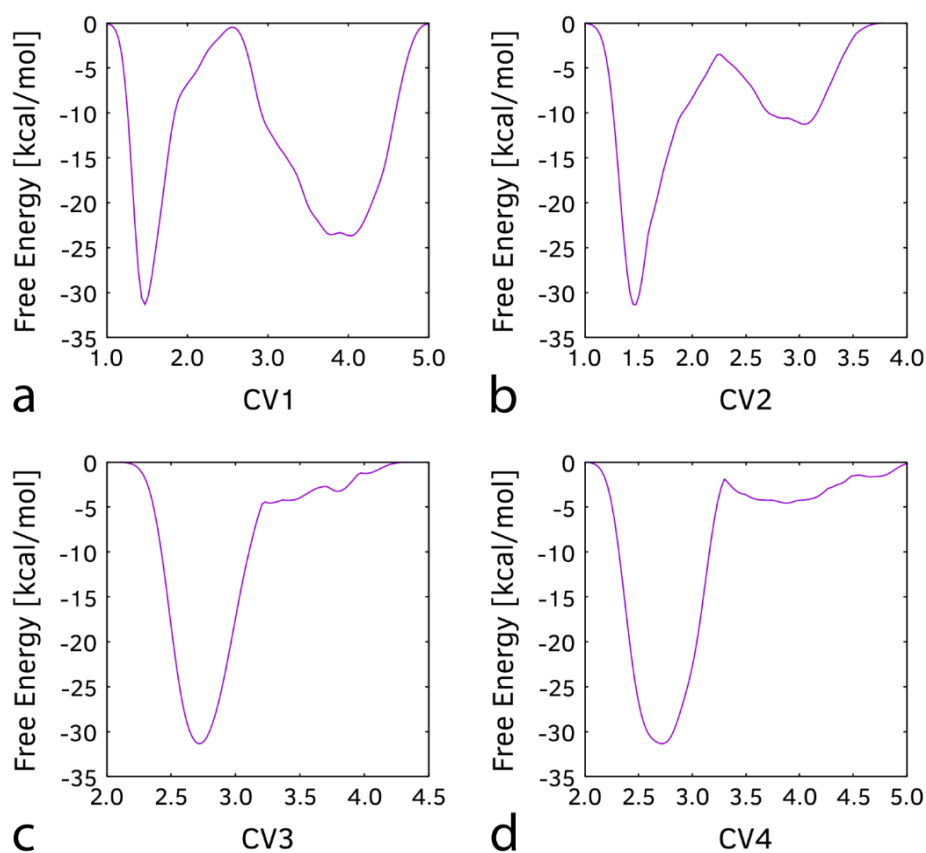

**Figure 12**: Free energy profiles of the gap-closing reactions for the transformation of C71SC to C72SC. The product minima are not completely explored.

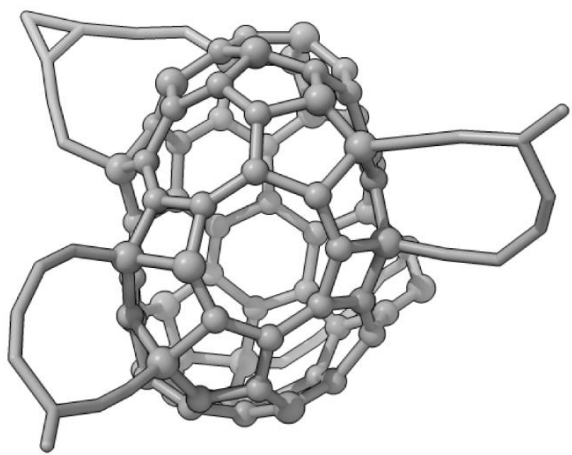

**Figure 13**. The geometry of C72SC – the final semi-spherical nanoparticle after gap-closing reactions.

*2.2 Modeling the synthesis of nanoparticle C66*

Another spherical nanoparticle is formed from CCII clusters with the use of metadynamics. The first step is the binding of two CCII fragments (Fig 14a). The fragment-coupling reactions are guided with three C – C distance CVs. All three reactions are almost barrierless (Fig 15). The small barriers found (< 2 kcal/mol) are once again probably an artifact of the time required for the fragments to bond at product bond lengths. A fourth C – C coupling reaction occurs without CV bias. All three set reactions are exothermic. The resulting cluster forms in less than 1 ps of simulation. Although the product free energy pit is not completely explored, according to the final energy profiles the product is an order of magnitude more stable than the reagents. The product is a half-sphere named 2×CCII (Fig 14b). This molecule is a single chain of 44 atoms in a bowl formation.

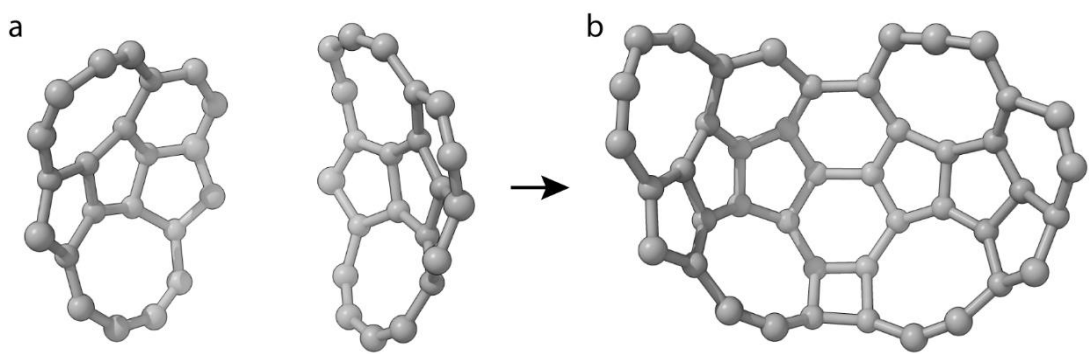

**Figure 14**. Geometries of the initial arrangement of the two CCII clusters (a) and the final structure (2×CCII) after DFTB metadynamics.

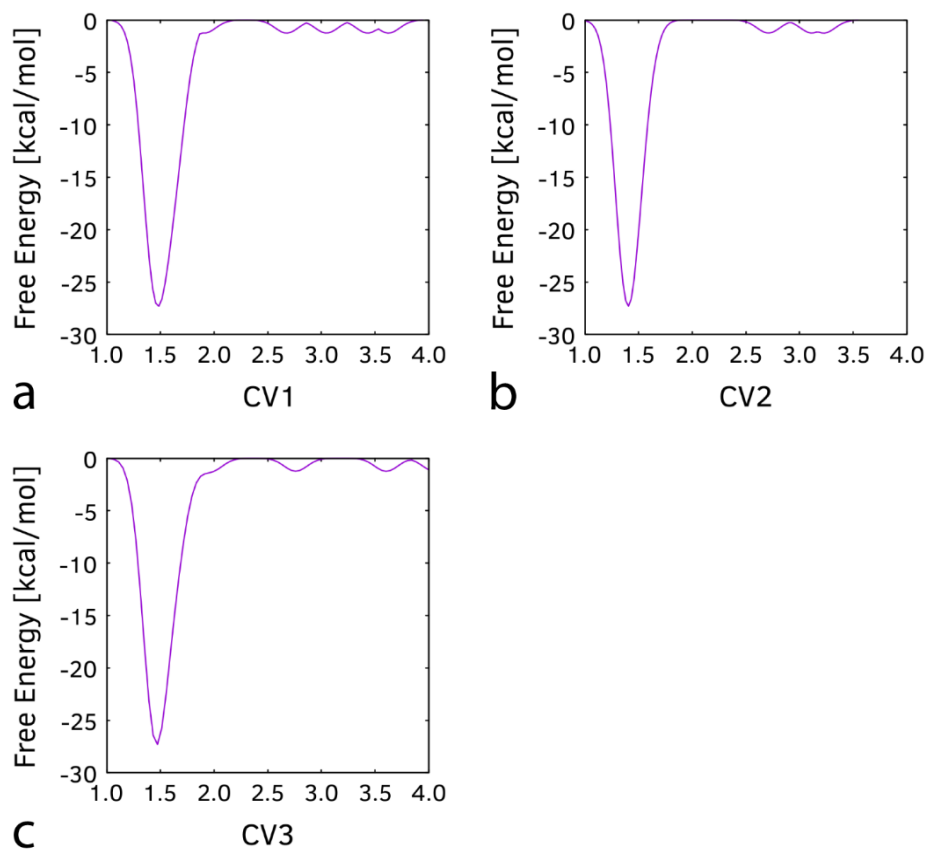

**Figure 15**: Free energy profiles of the guided 3 of the coupling reactions in the formation of 2×CCII. The product minima are not completely explored.

An additional CCII reacts with 2×CCII (Fig 16a) in another MTD simulation to a semi-spherical NP. The fragment-coupling reactions are guided with four C – C distance CVs. All four reactions are almost barrierless (Fig 17). A product forms in less than 1 ps. Further fragment-coupling bonds form without CV bias to a spherical structure (3×CCII, Fig 16b) with 66 atoms and no side chains. The barrier for the reverse process is as high as 27 kcal/mol manifesting the significant stability of the formed 3×CCII structure.

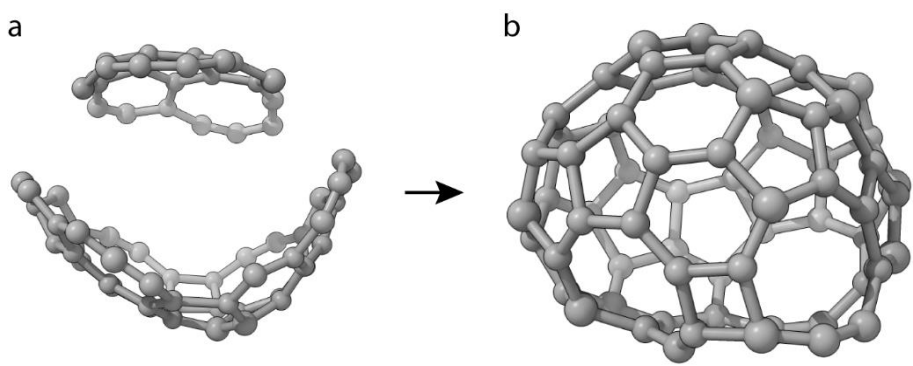

**Figure 16**: Geometries of the initial arrangement of the interacting CCII and 2×CCII clusters (a) and the final structure (3×CCII).

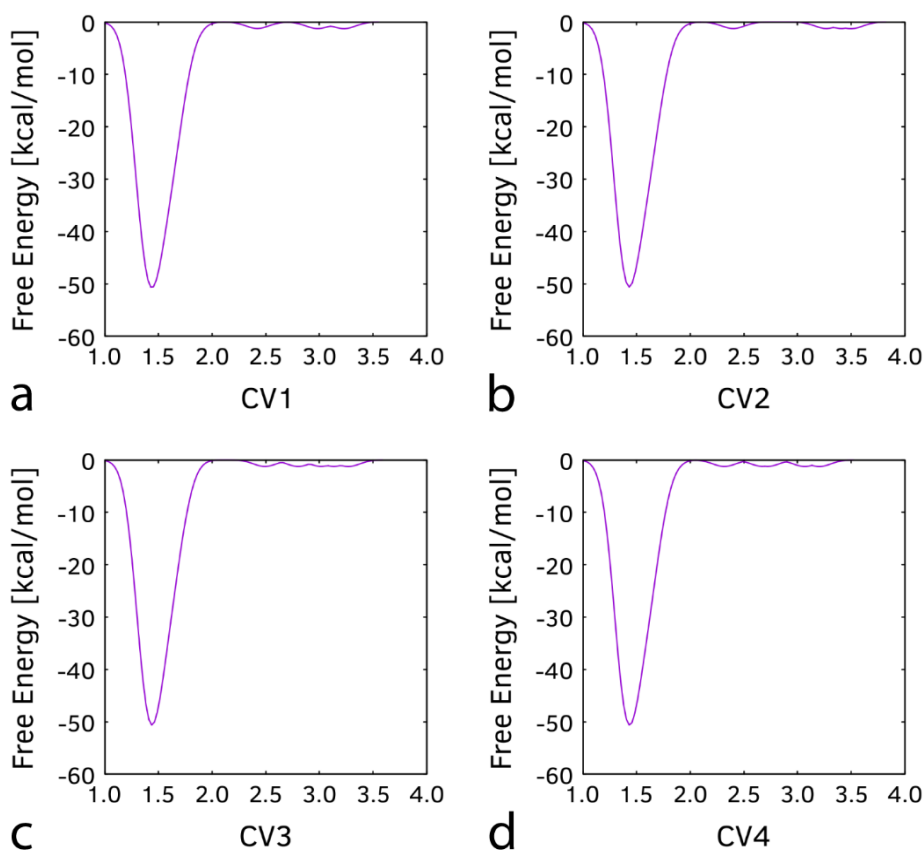

**Figure 17**: Free energy profiles of the guided four, among all of the coupling reactions in the formation of 3×CCII. The product minima are not completely explored.

*2.2.2 Structural improvements of 3xCCII*

Improvement in the structure of 3xCCII is carried out with MTD for a more symmetrical shape. A few gaps are to be closed by intramolecular bond formation. A total of three C – C distance CVs are used. The largest forward free energy barrier found is 4 kcal/mol (Fig 18). and gap-reopening reactions are very unlikely. The final fullerene-like nanoparticle is C66 and represents a C66[4,5,6,7,8,13]fullerene (Fig 19). Its binding energy per atom is the lowest among the studied molecules, at -8.26 eV. The cluster is not as symmetrical as fullerenes C60 or C70 but considering the fact that the initial state of the aggregation was carbon atoms scattered in a tetrahedral orientation to each other, the final result is significant.

In various "top-down" studies on fullerene synthesis, the pre-final step is of irregular, closed cage structures like C72SC and C66[66-68].

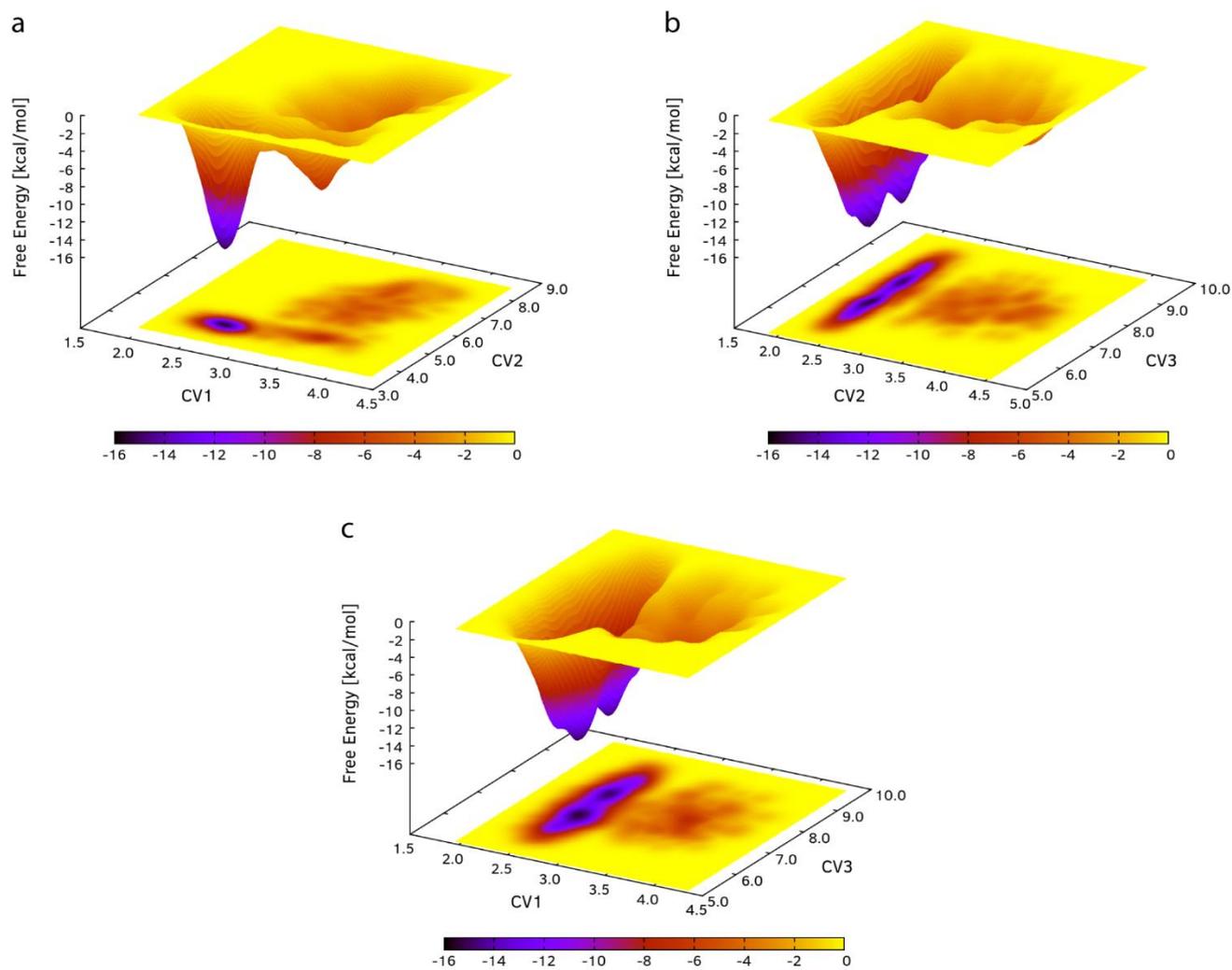

**Figure 18**: Free energy profiles of the gap-closing reactions for the transformation of 3×CCII to C66. The product minima are not completely explored.

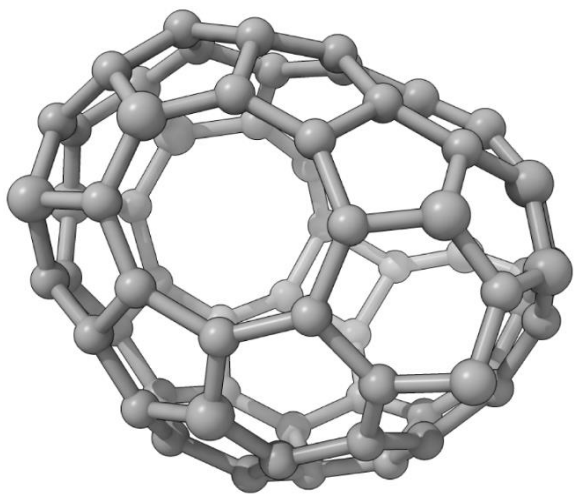

**Figure 19**. The geometry of C66 – the final semi-spherical nanoparticle after gap-closing reactions.

*3 IR-spectrum resemblance to known fullerenes*

The known fullerenes C60 and C70 are considered to have been detected in space[45,46], hence a comparison with the IR spectrum of the nanoparticles modeled in this article is in order. It is considered that UV radiation in space can break linear and circular carbon chains[91], hence of interest is also the form of C72SC without the side chains. The theoretical IR spectrums of C72, C72SC and C66 do resemble the theoretical spectrum of the fullerene C70, and C66 comes the closest (Fig 20). Studies with IR telescopes may detect such improperly shaped NPs and misclassify them as known fullerenes.

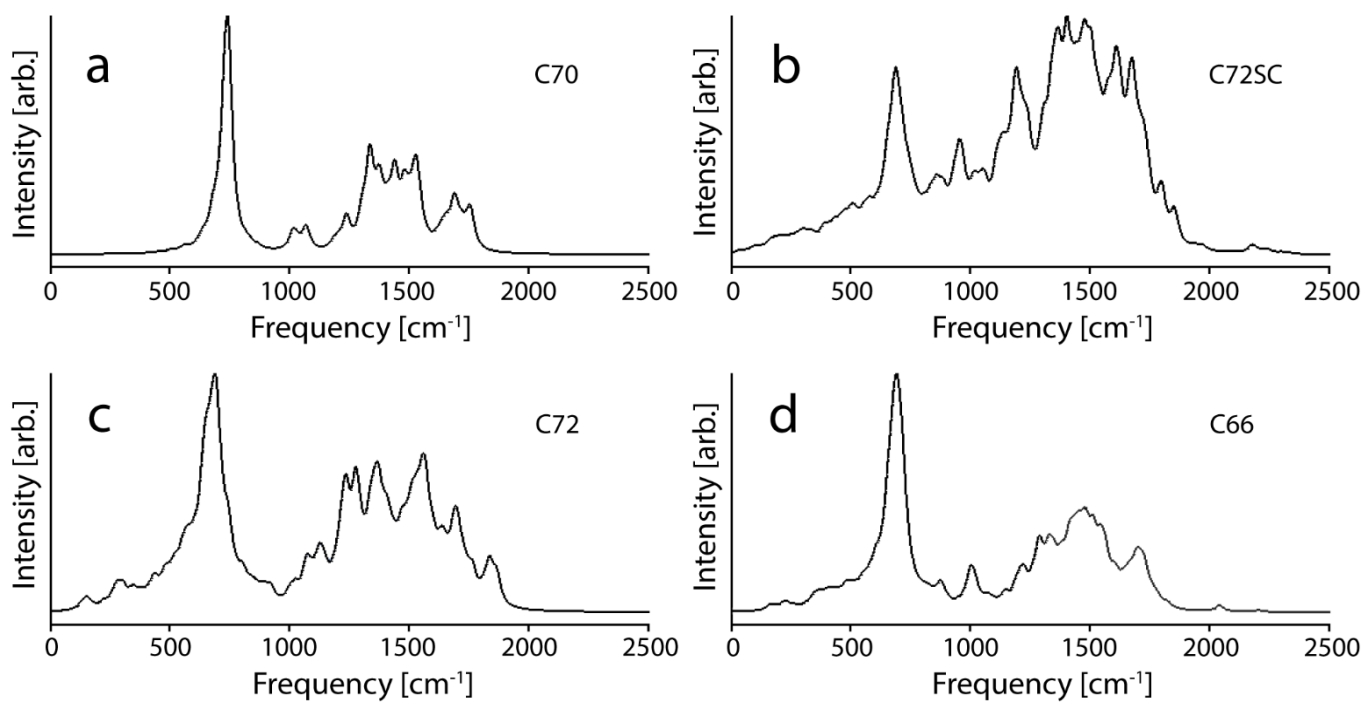

**Figure 20**. Theoretical IR spectrum of the optimized structures: (a) C70, (b) C72SC, (c) C72, and (d) C66.

## 4. Conclusions

Aggregates of carbon atoms with different initial orientations, densities, and periodicity were used as starting systems to study the formation of carbon clusters and nanophases in space. The stability and evolution of the species was tested with DFTB molecular dynamics. Spheroid carbon nanoparticles resembling imperfect fullerenes were modeled by coupling the resultant clusters through the use of metadynamics.

Isolated carbon aggregations of 20 to 25 atoms form $sp^2$-dominant clusters representing the C-skeleton of PAHs, given a sufficiently high original density. The results are very similar, regardless of initial interatomic orientation. Such clusters have a resemblance to fullerene building blocks. If the original density is sufficiently reduced, similar-sized aggregations result in branched molecules with similar amounts of C atoms in $sp$ and $sp^2$ hybridization and no carbons in $sp^3$ state. Such structures have very few rings of low atomic order. If the PBC model has varying cell dimensions, tailored by the cluster size, $sp^2$-dominant multidimensional polymers form at the higher initial density. Structural rearrangements are observed during DFTB dynamical simulations. Such simulations tend to show an increase in the quantity of atoms in $sp^2$ hybridization and a decrease in the quantity of $sp^3$ hybridized ones. Hence, we also expect such tendency to guide spontaneous transformations in nature. All 0-dimensional forms conform to experimentally detected types of molecules in space.

The spheroid nanoparticles were assembled with coupling reactions between primordial carbon clusters. One of those clusters was the result of the spontaneous chemistry between carbon atoms randomly arranged in their initial state. Another cluster formed spontaneously from carbon atoms with a tetrahedral arrangement in the initial system. In each interfragment coupling simulation most to all C – C binding reactions are barrierless. Structure C66 has the lowest binding energy per atom among the studies molecules. Occasionally, a misfit of atomic positions or a required slight whole-body twist may result in a single low barrier. The theoretical IR spectrum of two of the spheroid nanoparticles closely resembles that of fullerene C70 and it is possible that such imperfect species can be misidentified as known fullerenes in experimental IR studies of carbon signal sources in space.

**Acknowledgments**

The authors gratefully acknowledge financial support by the National Science Fund of Bulgaria under grant KP-06-COST/10 – 07.08.2023 in the framework of the COST Action CA21126 NanoSpace.
The authors also acknowledge the provided access to the e-infrastructure of the NCHDC – part of the Bulgarian National Roadmap on RIs, with financial support by Grant No D01-168/28.07.2022.